\documentclass[a4paper,12pt,reqno]{article}
\usepackage{amsmath,amsfonts,amssymb,amsthm,dsfont,slashed}
\allowdisplaybreaks
\newcommand{\da}{{\dot \alpha}}
\newcommand{\dbe}{{\dot \beta}}
\newcommand{\dga}{{\dot \gamma}}
\newcommand{\de}{\delta}

\newcommand{\be}{\beta}
\newcommand{\dR}{{(r)}}

\newcommand{\cg}{{\cal G}}
\newcommand{\cA}{{\cal A}}
\newcommand{\cB}{{\cal B}}
\newcommand{\cL}{{\cal L}}
\newcommand{\cT}{{\cal T}}
\newcommand{\qd}{{\quad}}
\newcommand{\LM}{M}

\newcommand{\TT}[3]{{T_{#1#2}}{}^{#3}}
\newcommand{\FF}[3]{{F_{#1#2}}{}^{#3}}
\newcommand{\f}[3]{{f_{#1#2}}{}^{#3}}
\newcommand{\Gg}[3]{{g_{#1#2}}{}^{#3}}

\newcommand{\spin}[2]{\omega_{#2}{}^{#1}}
\newcommand{\RS}[2]{\psi_{#2}{}^{#1}}
\newcommand{\bRS}[2]{\5\psi_{#2}{}^{#1}}
\newcommand{\viel}[2]{e_{#2}{}^{#1}}
\newcommand{\Viel}[2]{e^{#1}{}_{#2}}

\newcommand{\bea}{\begin{eqnarray}} \newcommand{\eea}{\end{eqnarray}}
\newcommand{\beann}{\begin{eqnarray*}} \newcommand{\eeann}{\end{eqnarray*}}
\newcommand{\beq}{\begin{equation}} \newcommand{\eeq}{\end{equation}}
\newcommand{\ba}{\begin{array}} \newcommand{\ea}{\end{array}}
\newcommand{\ben}{\begin{enumerate}} \newcommand{\een}{\end{enumerate}}

\newcommand{\om}{\omega}

\newcommand{\4 }{\tilde}
\newcommand{\5}{\bar}
\newcommand{\6}{\partial}

\newcommand{\DD}{{\cal D}}
\newcommand{\csum}[2]{\sum_{#1}\hspace{-1.#2em}\circ\ \ \ }

\newcommand{\mysection}[1]{\section{#1}
            \setcounter{equation}{0}\setcounter{figure}{0}}

\begin{document}

\begin{center}
 {\large\bfseries Properties of an alternative off-shell formulation of 4D supergravity}
 \\[5mm]
 Friedemann Brandt \\[2mm]
 \textit{Institut f\"ur Theoretische Physik, Leibniz Universit\"at Hannover, Appelstra\ss e 2, 30167 Hannover, Germany}
\end{center}

\begin{abstract}
This article elaborates on an off-shell formulation of D=4, N=1 supergravity whose auxiliary fields comprise an antisymmetric tensor field without gauge degrees of freedom. 
In particular, the relation to new minimal supergravity, a supercovariant tensor calculus and the construction of invariant actions including matter fields are discussed.
\end{abstract}

%\mysection{Introduction}\label{intro}

%\newpage
%\setcounter{tocdepth}{1}
%\tableofcontents

\mysection{Introduction}\label{intro}

In the basic formulation \cite{Freedman:1976xh,Deser:1976eh} of pure D=4, N=1 supergravity the commutator algebra of local symmetry transformations closes only on-shell. This complicates various computations, such as the construction of couplings of the supergravity multiplet (i.e. the supersymmetry multiplet with the vierbein and the gravitino) to matter multiplets, of locally supersymmetric invariants with higher derivatives and of Faddeev-Popov terms. Fortunately there are off-shell formulations of the theory with auxiliary fields that close the algebra of local symmetry transformations off-shell.

The best-known off-shell formulations of pure D=4, N=1 supergravity are the so-called old minimal formulation \cite{Stelle:1978ye,Ferrara:1978em} and the so-called new minimal formulation \cite{Sohnius:1981tp}. The auxiliary fields of the new minimal supergravity multiplet are a real 2-form gauge potential and a real vector field which is the gauge field of local R-transformations (``R-gauge field''). This auxiliary field content of new minimal supergravity hinders the algebraic elimination of the auxiliary fields because the equations of motion for these fields only contain the field strengths of the 2-form gauge potential and of the R-gauge field, i.e. derivatives of the auxiliary fields.

The subject of this article is an off-shell formulation \cite{Brandt:1997uq,Brandt:1997ny} of D=4, N=1 supergravity which overcomes this obstacle of new minimal supergravity. This formulation is a consistent deformation of new minimal supergravity coupled to an abelian gauge multiplet wherein the 2-form gauge potential mutates into an ordinary auxiliary field without gauge degrees of freedom and the R-gauge field mutates from an auxiliary field into a physical gauge field. The physical fields of the supergravity multiplet of this formulation are the vierbein, the R-gauge field, the gravitino and a spin-1/2 field, 
the auxiliary fields are a real antisymmetric two-component tensor field without gauge degrees of freedom and a real scalar field. This supergravity multiplet, according to the usual counting, has off-shell 16 bosonic degrees of freedom (6 from the vierbein, 3 from the R-gauge field, 6 from the auxiliary antisymmetric tensor field, 1 from the auxiliary scalar field) and 16 fermionic degrees of freedom (12 from the gravitino, 4 from the spin-1/2 field), and on-shell 4 bosonic degrees of freedom (from the vierbein and the R-gauge field) and 4 fermionic degrees of freedom (from the gravitino and the spin-1/2 field). Elimination of the auxiliary fields provides a supergravity model as in \cite{Freedman:1976uk} with gauged R-symmetry and spontaneously broken supersymmetry. A similar model has been found recently in \cite{Antoniadis:2020qoj}. 

In \cite{Brandt:1997uq,Brandt:1997ny} it was noted already that the coupling of the supergravity multiplet to other supersymmetry multiplets (``matter multiplets'') is analogous to the coupling of the new minimal supergravity multiplet to these multiplets. However, details were not given. The present paper is committed to providing these details and elaborating on related features of the theory. In section \ref{def} the model of \cite{Brandt:1997uq,Brandt:1997ny} is revisited. In section \ref{trafos} a supercovariant tensor calculus is presented and the symmetry transformations of the component fields of matter multiplets are given explicitly as nilpotent Becchi-Rouet-Stora-Tyutin (BRST) transformations. In section \ref{action} locally supersymmetric actions for the matter multiplets are constructed and the elimination of the auxiliary fields is discussed. Section \ref{con} contains a brief discussion of the results.

The conventions used here are the same as in \cite{Brandt:1996au} and differ from those of \cite{Wess:1992cp} basically only in the choice of the Minkowski metric which is $\eta_{ab}\sim (1,-1,-1,-1)$. In particular, spinors are Weyl spinors in the van der Waerden notation. Throughout the paper the component formalism of supergravity is used. Superfields and superspace techniques are not used at all. Instead, BRST techniques 
are used which are briefly recapitulated in an appendix.

 \mysection{Supergravity multiplet and its Lagrangian}\label{def}
 
 Our starting point is the deformation \cite{Brandt:1997uq,Brandt:1997ny} of new minimal supergravity coupled to an abelian gauge multiplet. The fields of that supergravity model are the components fields of the new minimal supergravity multiplet which are the vierbein ${e_\mu}^a$, the gravitino $\psi_\mu$, a real 2-form gauge potential with components $t_{\mu\nu}=-t_{\nu\mu}$ and a real R-gauge field $A^\dR_\mu$, and the component fields of the abelian gauge multiplet which are a real gauge field $A_\mu$, a complex spinor field $\lambda$ and an auxiliary real scalar field $D$. The Lagrangian ${\cal L}_{SG}$ for these fields derived in \cite{Brandt:1997uq,Brandt:1997ny} reads:
 \bea
 {\cal L}_{SG}&=&M_{Pl}^2{\cal L}_1+g_0^{-2}{\cal L}_2+g_2{\cal L}_3\label{Ldef}\\
{\cal L}_1/e&=&
\tfrac{1}{2} R
-2\varepsilon^{\mu\nu\rho\sigma}
(\psi_\mu\sigma_\nu\nabla_\rho\bar \psi_\sigma +c.c.)
-3H_\mu H^\mu
-2\varepsilon^{\mu\nu\rho\sigma}A^\dR_\mu \partial_\nu t_{\rho\sigma}
\label{Lgrav}\\
{\cal L}_2/e&=&
-\tfrac{1}{4}
(F_{\mu\nu}+g_1 t_{\mu\nu})(F^{\mu\nu}+g_1 t^{\mu\nu})
+\tfrac{1}{2} D{}^2
-\tfrac{1}{8} g_1{}^2
\nonumber\\
& &
+\tfrac{3}{2} \lambda\sigma^\mu\bar {\lambda}\, H_\mu
-\tfrac{1}{2}[ i  \lambda\sigma^\mu\nabla_\mu\bar {\lambda}
             +ig_1\psi_\mu\sigma^\mu\bar {\lambda}
\nonumber\\
& &
+\varepsilon^{\mu\nu\rho\sigma}(F_{\mu\nu}+g_1 t_{\mu\nu})\,
\psi_\rho\sigma_\sigma\bar{\lambda}
-2\psi_\mu\sigma^{\mu\nu}\psi_\nu\bar {\lambda}\bar {\lambda}+c.c.]
\label{LU1}
\\
{\cal L}_{3}/e&=&
D+\lambda\sigma^\mu\bar \psi_\mu+\psi_\mu\sigma^\mu\bar {\lambda}
+\varepsilon^{\mu\nu\rho\sigma}
(A_\mu \partial_\nu t_{\rho\sigma}
+\tfrac{1}{4}g_1 t_{\mu\nu} t_{\rho\sigma})
\label{LFI}
\eea
where
\bea
e&=&\det (\viel a\mu)\label{e}\\
\varepsilon^{\mu\nu\rho\sigma}&=&\Viel \mu a\Viel \nu b\Viel \rho c\Viel \sigma d
\varepsilon^{abcd}\label{epsilon}\\
R&=&2\Viel \nu a\Viel \mu b
(\partial_{[\mu}{\omega_{\nu]}}^{ab}
-{\omega_{[\mu}}^{ca}{\omega_{\nu]c}}^b)
\label{Riemann}\\
H^\mu&=&\varepsilon^{\mu\nu\rho\sigma}
        (\tfrac{1}{2} \partial_{\nu}t_{\rho\sigma}
         +i \psi_{\nu}\sigma_\rho\bar \psi_{\sigma})
\label{H}\\
F_{\mu\nu}&=&2(\partial_{[\mu} A_{\nu]}
     +i \lambda\sigma_{[\mu}\bar \psi_{\nu]}
               +i \psi_{[\mu}\sigma_{\nu]}\bar {\lambda}) 
\label{F}\\
\nabla_\mu\psi_\nu&=&\partial_\mu\psi_\nu
-\tfrac{1}{2}{\omega_\mu}^{ab}\psi_\nu\sigma_{ab}
-i A^\dR_\mu\psi_\nu
\label{nablapsi}\\
\nabla_\mu\lambda&=&
\partial_\mu\lambda
-\tfrac{1}{2}{\omega_\mu}^{ab}\lambda\sigma_{ab}
-i A^\dR_\mu\lambda
\label{nablala}\\
\spin {ab}\mu &=& e^{\nu a}e^{\rho b}(\omega_{[\mu\nu]\rho}
-\omega_{[\nu\rho]\mu}+\omega_{[\rho\mu]\nu})
\label{spin1}\\
\omega_{[\mu\nu]\rho}&=&e_{\rho a}\6_{[\mu}\viel a{\nu]}
-i\psi_{\mu}\sigma_\rho\5\psi_{\nu}
+i\psi_{\nu}\sigma_\rho\5\psi_{\mu}.
\label{spin2}
\eea
$M_{Pl}$, $g_0$, $g_1$ und  $g_2$ are real coupling constants. $ \Viel \mu a$ denotes the inverse vierbein fulfilling 
\[ \Viel \mu a\viel a\nu =\delta^\mu_\nu,\quad \Viel \mu b\viel a\mu =\delta^a_b.\]
The three contributions ${\cal L}_1$, ${\cal L}_2$ and ${\cal L}_3$ to the Lagrangian (\ref{Ldef}) are separately invariant up to a total divergence, respectively, under general coordinate transformations, local supersymmetry transformations, local Lorentz transformations, local R-transformations and local gauge transformations of $t_{\mu\nu}$ and $A_\mu$. The local supersymmetry transformations and the gauge transformations of $t_{\mu\nu}$ and $A_\mu$ are, written as part $\hat s$ of the BRST transformations of these fields:
\bea
 \hat s\viel a\mu &=& 
2i(\xi\sigma^a\5\psi_\mu-\psi_\mu\sigma^a\5\xi)
\label{4sviel}\\
\hat s\psi_\mu&=&\6_\mu\xi-\tfrac{1}{2}\spin {ab}\mu\xi\sigma_{ab}-i A^\dR_\mu\xi
-i \xi H_\mu-i \xi\sigma_{\mu\nu} H^\nu
\label{4spsi}\\
\hat st_{\mu\nu}&=&2\6_{[\nu} Q_{\mu]}-2i(\xi\sigma_{[\mu}\5\psi_{\nu]}
+\psi_{[\mu}\sigma_{\nu]}\5\xi)
\label{4st}\\
\hat sA^\dR_\mu&=&\xi\sigma_\mu \5S+S\sigma_\mu \5\xi
\label{4sa}\\
\hat sA_\mu&=&\6_\mu C
-i \xi\sigma_\mu\5\lambda
+i\lambda\sigma_\mu\5\xi+g_1 Q_\mu
\label{4sA}\\
\hat s\lambda&=&\xi (\tfrac{1}{2} g_1-iD)
-\xi\sigma^{\mu\nu}(F_{\mu\nu}+g_1 t_{\mu\nu})
\label{4slambda}\\
\hat sD&=&\xi \sigma^\mu \left[\nabla_\mu\5\lambda-\5\psi_\mu (iD+\tfrac{1}{2} g_1)
-\5\sigma^{\nu\rho}\5\psi_\mu (F_{\nu\rho}+g_1 t_{\nu\rho})
+\tfrac{3i}{2} \5\lambda H_\mu
\right]+c.c.\quad
\label{4sD}\eea
where $S$ in (\ref{4sa})
is the spin-$\frac 12$ portion of the supercovariant
gravitino field strength,
\beq
S=2(\nabla_\mu\psi_\nu)\sigma^{\mu\nu}
+\tfrac{3i}{2} \psi_\mu H^\mu ,\quad
\5S=-2\5\sigma^{\mu\nu}\nabla_\mu\5\psi_\nu
-\tfrac{3i}{2} \5\psi_\mu H^\mu .
\label{S}\eeq
$\xi^\alpha$ are ghosts of local supersymmetry transformations, $Q_\mu$ are ghosts of reducible gauge transformations of $t_{\mu\nu}$ and $C$ is a ghost of local gauge transformations of $A_\mu$. 

The Lagrangian (\ref{Ldef}) and the symmetry transformations given above are deformations of the Lagrangian and symmetry transformations of new minimal supergravity coupled to $A_\mu$,
$\lambda$ and $D$ with deformation parameters
$g_1$ and  
$g_2$. Only the symmetry transformations of $A_\mu$, 
$\lambda$ and $D$ are deformed by the $g_1$-dependent terms in (\ref{4sA}), (\ref{4slambda}) and (\ref{4sD}).  

Now, a first observation is that the dependence on $g_1$ of the symmetry transformations can be completely removed by the following rescalings of fields:
\bea
A^\prime_\mu=g_1^{-1}A_\mu,\ \lambda^\prime=g_1^{-1}\lambda,\ 
D^\prime=g_1^{-1}D,\ C^\prime=g_1^{-1}C.\label{fieldredef1}
\eea
A second observation is that for $g_1\neq 0$ the Lagrangian (\ref{Ldef}), up to a total divergence, and the symmetry transformations depend on 
$t_{\mu\nu}$ and $A_\mu$ only via the combination
\bea
b_{\mu\nu}=t_{\mu\nu}+\6_\mu A^\prime_\nu-\6_\nu A^\prime_\mu\label{fieldredef2}
\eea
because $A^\prime_\mu$ does not contribute to $H^\mu$ when $H^\mu$ is written in terms of $b_{\mu\nu}$ and the terms in ${\cal L}_{3}$ depending on $t_{\mu\nu}$ are equal to 
$\frac 14g_1 e\varepsilon^{\mu\nu\rho\sigma}b_{\mu\nu} b_{\rho\sigma}$ up to a total divergence. Redefining also the deformation parameters as
\bea
g^\prime_1=(g_1/g_0)^2,\quad g^\prime_2=g_1 g_2
\eea
the Lagrangian (\ref{Ldef}) can for $g_1\neq 0$ be written, up to a total divergence, as
\bea 
 {\cal L}_{SG}&=&M_{Pl}^2{\cal L}_1+g^\prime_1{\cal L}^\prime_2+g^\prime_2{\cal L}^\prime_3\label{LSG}\\
{\cal L}_1/e&=&
\tfrac{1}{2} R
-2\varepsilon^{\mu\nu\rho\sigma}
(\psi_\mu\sigma_\nu\nabla_\rho\bar \psi_\sigma +c.c.)
-3H_\mu H^\mu
-2\varepsilon^{\mu\nu\rho\sigma}A^\dR_\mu \partial_\nu b_{\rho\sigma}
\label{L1}\\
{\cal L}^\prime_2/e&=&
-\tfrac{1}{4} b_{\mu\nu}b^{\mu\nu}
+\tfrac{1}{2} D^\prime{}^2
-\tfrac{1}{8}
+\tfrac{3}{2}\lambda^\prime\sigma^\mu\bar {\lambda}^\prime H_\mu
+2(\lambda^\prime\sigma_{[\mu}\bar\psi_{\nu]})(\psi^\mu\sigma^\nu\bar\lambda^\prime)
\nonumber\\
&&-\tfrac{1}{2}( i  \lambda^\prime\sigma^\mu\nabla_\mu\bar {\lambda}^\prime
             +i\psi_\mu\sigma^\mu\bar {\lambda}^\prime
             +2ib^{\mu\nu}
\psi_\mu\sigma_\nu\bar{\lambda}^\prime
+\varepsilon^{\mu\nu\rho\sigma} b_{\mu\nu}\,
\psi_\rho\sigma_\sigma\bar{\lambda}^\prime
\nonumber\\
&&
+\psi_\mu\sigma^{\mu\nu}\psi_\nu\bar {\lambda}^\prime\bar {\lambda}^\prime
-\tfrac 32 \psi_\mu\psi^\mu\bar {\lambda}^\prime\bar {\lambda}^\prime
+c.c.)
\label{L2}
\\
{\cal L}^\prime_{3}/e&=&
D^\prime+\lambda^\prime\sigma^\mu\bar \psi_\mu+\psi_\mu\sigma^\mu\bar {\lambda}^\prime
+\tfrac{1}{4}\varepsilon^{\mu\nu\rho\sigma}
b_{\mu\nu} b_{\rho\sigma}
\label{L3}
\eea
with $H^\mu$ now defined in terms of $b_{\mu\nu}$ according to
\beq
H^\mu=\varepsilon^{\mu\nu\rho\sigma}(\tfrac{1}{2}\6_{\nu}b_{\rho\sigma}
+i\psi_{\nu}\sigma_\rho\5\psi_{\sigma}).
\label{Hb}
\eeq
The fields in the Lagrangian (\ref{LSG}) are the vierbein ${e_\mu}^a$, the gravitino $\psi_\mu$, the R-gauge field $A^\dR_\mu$, the spin-1/2 field $\lambda^\prime$ and the auxiliary fields $b_{\mu\nu}$ and $D^\prime$. The complete BRST transformations of these fields are:
\bea
 s\viel a\mu &=& C^\nu\6_\nu\viel a\mu+(\6_\mu C^\nu)\viel a\nu
+C_b{}^a\viel b\mu
+2i(\xi\sigma^a\5\psi_\mu-\psi_\mu\sigma^a\5\xi)
\label{sviel}\\
s\psi_\mu&=&C^\nu\6_\nu\psi_\mu+(\6_\mu C^\nu)\psi_\nu
+\tfrac{1}{2} C^{ab}\psi_\mu\sigma_{ab}+iC^\dR  \psi_\mu
\nonumber\\
& &+\6_\mu\xi-\tfrac{1}{2}\spin {ab}\mu\xi\sigma_{ab}-i A^\dR_\mu\xi
-i \xi H_\mu-i \xi\sigma_{\mu\nu} H^\nu
\label{spsi}\\
sA^\dR_\mu&=&C^\nu\6_\nu A^\dR_\mu+(\6_\mu C^\nu)A^\dR_\nu
+\6_\mu C^\dR
+\xi\sigma_\mu \5S+S\sigma_\mu \5\xi
\label{sa}\\
s\lambda^\prime&=&C^\mu\6_\mu\lambda^\prime+\tfrac{1}{2} C^{ab}\lambda^\prime\sigma_{ab}
+i C^\dR\lambda^\prime
+\xi (\tfrac{1}{2} -iD^\prime)
-\xi\sigma^{\mu\nu}b_{\mu\nu}
\label{sl}\\
sb_{\mu\nu}&=&C^\rho\6_\rho b_{\mu\nu}
+2b_{\rho[\nu}\6_{\mu]} C^\rho
-2i\left(\6_{[\mu}[\xi\sigma_{\nu]}\5\lambda^\prime]
+\xi\sigma_{[\mu}\5\psi_{\nu]}-c.c.\right)
\label{sb}\\
sD^\prime&=&C^\mu\6_\mu D^\prime
+\left(\xi \sigma^\mu [\nabla_\mu\5\lambda^\prime-\5\psi_\mu (\tfrac{1}{2}+iD^\prime)
-\5\sigma^{\nu\rho}\5\psi_\mu b_{\nu\rho}
+\tfrac{3i}{2} \5\lambda^\prime H_\mu]+c.c.\right)\qquad
\label{sd}
\eea
where $S$ and $\5S$ are as in (\ref{S}) with $H^\mu$ as in (\ref{Hb}).
$C^\mu$ are ghosts of general coordinate transformations, $C^{ab}=-C^{ba}$ are ghosts of local Lorentz transformations and $C^\dR$ is a ghost of local R-transformations. The ghosts $C^\mu$, $C^{ab}$ and $C^\dR$ are real and 
Gra\ss mann odd, the supersymmetry ghosts $\xi^\alpha$ are complex and Gra\ss mann even, with $\bar\xi^\da$ denoting the complex conjugate of $\xi^\alpha$.
The BRST transformations of the ghosts are
\bea
sC^\mu&=&C^\nu\6_\nu C^\mu+2i\xi\sigma^\mu\5\xi\
\label{sCdiff}\\
s\xi&=&C^\mu\6_\mu \xi
+\tfrac{1}{2}C^{ab}\xi\sigma_{ab}
+iC^\dR\xi
-2i \xi\sigma^\mu\5\xi\,\psi_\mu\label{sxi}\\
sC^{ab}&=&C^\mu\6_\mu C^{ab}
+C^{ca}{C_c}^b-2i \xi\sigma^\mu\5\xi\,\spin {ab}\mu
+2i\varepsilon^{abcd}\xi\sigma_c\5\xi \, H_d
\label{sCLor}\\
sC^\dR&=&C^\mu\6_\mu C^\dR 
-2i \xi\sigma^\mu\5\xi\, A^\dR_\mu .
\label{sc}
\eea
In the BRST transformations (\ref{sviel}) through (\ref{sc}) all spinor fields have upper spinor indices.
These transformations are strictly nilpotent off-shell ($s^2=0$), i.e. the algebra of the corresponding local symmetry transformations closes off-shell. The three portions ${\cal L}_1$, ${\cal L}^\prime_2$, ${\cal L}^\prime_3$ of the Lagrangian (\ref{LSG}) are separately invariant up to a total divergence, respectively,  under the BRST transformations (\ref{sviel}) through (\ref{sd}) as these portions arise from ${\cal L}_1$, ${\cal L}_2$, ${\cal L}_3$, respectively. The gauge field $A_\mu$ and the ghosts $Q_\mu$ and $C$ have completely disappeared from the theory, along with the corresponding gauge symmetries.
We shall use this formulation in the following analysis. Of course one can return to the formulation
with Lagrangian (\ref{Ldef}) by undoing the field redefinitions (\ref{fieldredef1}) and (\ref{fieldredef2}).

The Lagrangian (\ref{LSG}) is quite similar to the one given in equations (4.16) and (4.17) of \cite{Antoniadis:2020qoj}. 
Apart from different conventions, (\ref{LSG}) differs from equations (4.16) and (4.17) of \cite{Antoniadis:2020qoj} 
in the use of 
$b_{\mu\nu}$ instead of its supercovariant counterpart $B_{ab}$ given below in equation (\ref{B}). Furthermore
(\ref{LSG}) differs from equations (4.16) and (4.17) of \cite{Antoniadis:2020qoj} by the term $\tfrac{3}{2}\lambda^\prime\sigma^\mu\bar {\lambda}^\prime H_\mu$ in (\ref{L2}) (and possibly by some 4-fermion terms which is hard to check). The term
$\tfrac{3}{2}\lambda^\prime\sigma^\mu\bar {\lambda}^\prime H_\mu$ is needed in order that (\ref{L2}) is invariant off-shell
up to a total divergence under the BRST transformations (\ref{sviel}) through (\ref{sd}) and is present already in the undeformed model (i.e. for $g_1=0$) -- in fact such a term occurs also in equations (5.8) and (5.9) of \cite{Sohnius:1982fw} where it is present in $i\bar\lambda\hat{\slashed D}\lambda$, cf. equations (2.2) and (2.1) of \cite{Sohnius:1982fw}. It is an open issue whether this difference
between (\ref{LSG}) and equations (4.16) and (4.17) of \cite{Antoniadis:2020qoj} is significant or resolvable, for instance by field redefinitions.\footnote{It is not evident whether or not such field redefinitions exist. For instance, a redefinition 
$\tilde A_\mu^\dR=A_\mu^\dR-\tfrac 38 M_{Pl}^{-2}g_1^{\prime}
\lambda^\prime\sigma_\mu\bar {\lambda}^\prime$ of the $R$-gauge field removes the term $\tfrac{3}{2}\lambda^\prime\sigma^\mu\bar {\lambda}^\prime H_\mu$ from (\ref{LSG}) but introduces a 4-fermion coupling proportional to 
$\lambda^\prime\lambda^\prime\bar {\lambda}^\prime\bar {\lambda}^\prime$ which apparently has no counterpart in equations (4.16) and (4.17) of \cite{Antoniadis:2020qoj}.}
 
\mysection{Matter multiplets and supercovariant tensor calculus}\label{trafos}

In order to couple matter multiplets to the supergravity multiplet and to construct supersymmetric actions for these multiplets we use a supercovariant tensor calculus. The calculus comprises supercovariant derivatives $\DD_a$, spinorial anti-derivations $\DD_\alpha$, $\bar\DD_\da$ and generators $\de_I$ of a structure group which are realized on supercovariant tensors (see below) and fulfill the graded commutator algebra
\beq [\DD_A,\DD_B\}=-\TT ABC\DD_C-\FF ABI\de_I,\qd
[\de_I,\DD_A]=-\Gg IAB\DD_B,\qd
[\de_I,\de_J]=\f IJK\de_K                                      \label{s5}\eeq
where the index $A$ of $\DD_A$ runs over Lorentz vector indices $a$ and spinor indices $\alpha, \da$. 
$[\DD_A,\DD_B\}$ denotes the commutator $[\DD_A,\DD_B]$ if $A$ or $B$ is a Lorentz vector index and the anticommutator $\{\DD_A,\DD_B\}$ if both $A$ and $B$ are spinor indices. The $\f IJK$ denote structure constants of 
the Lie algebra $\cg$ of the structure group which is the direct sum of the Lorentz algebra and a further reductive
Lie algebra which at least comprises the generator $\de_\dR$ of R-transformations and may comprise further generators 
$\delta_i$ of a Yang-Mills gauge group with or without abelian factors. Denoting the generators of the Lorentz algebra by 
$l_{ab}=-l_{ba}$, we have
\[ \{\DD_A\}=\{\DD_a,\DD_\alpha,\5\DD_\da\},\qd \{\de_I\}=\{l_{ab},\de_\dR,\delta_i\}.
\]
The sum over indices of $\cg$ is defined with a factor 1/2 for the Lorentz generators, such as
\[
\FF ABI\de_I=\tfrac 12\FF AB{ab}l_{ab}+\FF AB\dR\de_\dR+\FF ABi\de_i
\]
and the sum over indices $A,B,\ldots$ is defined with upper first spinor index, such as
\[
\TT ABC\DD_C=\TT ABc\DD_c+\TT AB\gamma\DD_\gamma+\TT AB\dga\bar\DD_\dga .
\]
$\Gg IAB$ are the entries of a matrix $g_I$ which represents $\de_I$
on the $\DD_A$. The only nonvanishing $\Gg IAB$ occur for $\de_I\in\{l_{ab},\de_\dR\}$ with
\bea 
[l_{ab},\DD_c]=\eta_{bc}\DD_a-\eta_{ac}\DD_b,&
[l_{ab},\DD_\alpha]=-\sigma_{ab\alpha}{}^\be\DD_\be,&
[l_{ab},\5\DD_\da]=\bar\sigma_{ab}{}^\dbe{}_\da\5\DD_\dbe\\ \mbox{}
[\de_\dR,\DD_a]=0,&
[\de_\dR,\DD_\alpha]=-i\DD_\alpha,&
[\de_\dR,\5\DD_\da]=i \5\DD_\da .
\label{s7}\eea

The matter multiplets treated here are chiral multiplets \cite{Wess:1974tw}, super-Yang-Mills multiplets (in WZ gauge)
\cite{Wess:1974jb,Ferrara:1974pu,Salam:1974ig} and linear multiplets \cite{Siegel:1979ai}. 
The component fields of the chiral multiplets are denoted by $\phi^m$, $\chi_\alpha^m$, $F^m$ and their complex conjugates $\bar\phi^{\bar m}$, $\bar \chi_{\dot\alpha}^{\bar m}$, $\bar F^{\bar m}$ where $\phi^m$, $F^m$ are complex scalar fields and $\chi^m$ are complex spinor fields. The component fields of the super-Yang-Mills multiplets 
are denoted by $A_\mu^i$, $\lambda^i_\alpha$,
$\bar \lambda_{\dot\alpha}^i$, $D^i$ where $A_\mu^i$ are real gauge fields, $D^i$ are real scalar fields, $\lambda^i$ are complex spinor fields and $\bar \lambda^i$ is the complex conjugate of $\lambda^i$. The component fields of the linear multiplets are denoted by $\varphi^{\LM}$, $A_{\mu\nu}^{\LM}$, $\psi_\alpha^{\LM}$, $\bar\psi_{\dot\alpha}^{\LM}$ where $\varphi^{\LM}$ are real scalar fields, $A_{\mu\nu}^{\LM}=-A_{\nu\mu}^{\LM}$ are real components of 2-form gauge potentials, $\psi^{\LM}$ are complex spinor fields and 
$\bar\psi^{\LM}$ is the complex conjugate of $\psi^{\LM}$. 

The supercovariant tensors which the supercovariant algebra (\ref{s5}) is realized on are 
$\phi^m$, $\chi^m$, $F^m$, 
$\bar\phi^{\bar m}$, $\bar \chi^{\bar m}$, $\bar F^{\bar m}$, 
$\lambda^i$, $\bar \lambda^i$, $D^i$,
$\varphi^{\LM}$, $\psi^{\LM}$, $\bar\psi^{\LM}$, 
$\lambda^\prime$, $\bar\lambda^\prime$, $B_{ab}$, $D^\prime$, $H^a=\viel a\mu H^\mu$,
$\TT ab\alpha$, ${\bar T}_{ab}{}^\da$, $\FF abI$,  $L^{\LM}_a$ 
and supercovariant derivatives of these tensors, with $H^\mu$ as in (\ref{Hb}) and 
$B_{ab}$, $\TT ab\alpha$, $\bar T_{ab}{}^\da$, $\FF abI$ and $L^{\LM}_a$ given by:
\bea 
B_{ab}&=&\Viel \mu a\Viel \nu b(b_{\mu\nu}+2i\lambda^\prime\sigma_{[\mu}\bar\psi_{\nu]}+2i\psi_{[\mu}\sigma_{\nu]}\bar\lambda^\prime)\label{B}\\
\TT ab\alpha&=&2\Viel \mu a\Viel \nu b(\nabla_{[\mu}\psi_{\nu]}
-iH_{[\mu} \psi_{\nu]}+iH^\rho\psi_{[\mu}\sigma_{\nu]\rho} )^\alpha
\label{Tab}\\
{\bar T}_{ab}{}^\da&=& 2\Viel \mu a\Viel \nu b(\nabla_{[\mu}\bar\psi_{\nu]}
+iH_{[\mu} \bar\psi_{\nu]}+iH^\rho\bar\sigma_{\rho[\mu}\bar\psi_{\nu]} )^\da
\label{barTab}\\
\FF ab{cd}&=&2\Viel \mu a\Viel \nu b[\6_{[\mu}\omega_{\nu]}{}^{cd}-\omega_{[\mu}{}^{ec}\omega_{\nu]e}{}^d
+i\psi_{[\mu}(\sigma_{\nu]}\bar T^{cd}-2\sigma^{[c}\bar T^{d]}{}_{\nu]})\nonumber\\
&&+i(T^{cd}\sigma_{[\mu}+2T^{[c}{}_{[\mu}\sigma^{d]})\bar\psi_{\nu]}
-2i\psi_{[\mu}\sigma^e\bar\psi_{\nu]}\varepsilon^{cd}{}_{ef}H^f]\label{Fabcd}\\
\FF ab\dR&=&2\Viel \mu a\Viel \nu b(\6_{[\mu}A^\dR_{\nu]}-\psi_{[\mu}\sigma_{\nu]}\bar S
+S\sigma_{[\mu}\bar\psi_{\nu]})\label{Fabr}\\
\FF abi&=&2\Viel \mu a\Viel \nu b(\6_{[\mu}A_{\nu]}^i
+\tfrac 12\f jki A_{\mu}^j A_{\nu}^k
+i\psi_{[\mu}\sigma_{\nu]}\bar \lambda^i
+i \lambda^i\sigma_{[\mu}\bar\psi_{\nu]})\label{Fabi}\\
L^{\LM}_a&=&e_{\mu a}\varepsilon^{\mu\nu\rho\sigma}(\tfrac 12\6_\nu A_{\rho\sigma}^\LM
-\psi_\nu\sigma_{\rho\sigma}\psi^\LM
+\bar\psi^\LM\bar\sigma_{\rho\sigma}\bar\psi_\nu
-2i\varphi^\LM\psi_\nu\sigma_\rho\bar\psi_\sigma)+2\varphi^\LM H_a\nonumber\\
&=&e_{\mu a}\varepsilon^{\mu\nu\rho\sigma}(\tfrac 12\6_\nu A_{\rho\sigma}^\LM
+\varphi^\LM\6_\nu b_{\rho\sigma}-\psi_\nu\sigma_{\rho\sigma}\psi^\LM
+\bar\psi^\LM\bar\sigma_{\rho\sigma}\bar\psi_\nu).\label{La}
\eea

$\delta_\dR$ is represented on supercovariant tensors according to table 1 where $r^m$ are real constants (``R-charges'' of the $\phi^m$). For the respective 
complex conjugate supercovariant tensors 
we have $\delta_\dR\bar{\Phi}=\overline{\delta_\dR\Phi}$ where $\bar{\Phi}$ denotes the complex conjugate of $\Phi$, and
$\overline{\delta_\dR\Phi}$ denotes the complex conjugate of $\delta_\dR\Phi$.
Real supercovariant tensors, such as $D^i$, $\varphi^{\LM}$, $B_{ab}$, $D^\prime$, $\FF abI$ and  $L^{\LM}_a$, 
have vanishing $R$-transformation.
\[
\ba{c|c|c|c|c|c|c|c}
\Phi                &\phi^m &\chi^m &F^m & \lambda^i & 
\psi^{\LM} & \lambda^\prime & 
\TT ab\alpha \\
\hline\rule{0em}{2.5ex}
\delta_\dR\Phi &ir^m\phi^m & i(r^m-1)\chi^m & i(r^m-2)F^m & i\lambda^i & 
-i\psi^\LM & i\lambda^\prime & i\TT ab\alpha \\
\multicolumn{8}{c}{}\\
\multicolumn{8}{c}{\mbox{Table 1}}
\ea
\]

$\DD_\alpha$ is realized on supercovariant tensors according to table 2
with (in the last row) $\TT \alpha a\beta$, $\FF \alpha\da{cd}$ and  $\FF a\alpha{cd}$ as in equations (\ref{TF2}), (\ref{TF4}) and (\ref{TF5}), respectively.
\[
\ba{c|l}
\Phi             & \DD_\alpha\Phi \\
\hline\rule{0em}{2ex}
\phi^m & \chi^m_\alpha\\
\chi^m_\beta  &  \varepsilon_{\beta\alpha}F^m \\
 F^m&0 \\
\bar\phi^m & 0\\
\bar\chi^m_\da & -2i\DD_{\alpha\da}\bar\phi^m\\
\bar F^m & (2i\DD_{\alpha\da}+H_{\alpha\da})\bar\chi^{m \da}
                                   -4(\lambda_\alpha^i\delta_i-i S_\alpha\delta_\dR)\bar\phi^m \\
\lambda^{i\beta}  &  - i  \delta_\alpha^\beta D^i-\sigma^{ab}{}_\alpha{}^\beta \FF abi \\
\bar\lambda^{i\da} & 0  \\
 D^i  & (\DD_{\alpha\da}+\tfrac {3i}2H_{\alpha\da})\bar\lambda^{i\da} \quad \\
 \FF abi&-2i\DD_{[a}\bar\lambda^{i\da} \sigma_{b]\alpha\da}
+H_{[a}\sigma_{b]\alpha\da}\bar\lambda^{i\da}
-i\varepsilon_{abcd}H^c\sigma^d{}_{\alpha\da}\bar\lambda^{i\da} \\
\varphi^{\LM} & \psi^{\LM}_\alpha \\
\psi^{\LM}_\beta & 0\\
 \bar\psi^{\LM}_\da &-i \DD_{\alpha\da}\varphi^{\LM}-L^\LM_{\alpha\da} \\
 L^{\LM}_a & 2i\sigma_{ab\alpha}{}^\beta\DD^b\psi^{\LM}_\beta+\tfrac 12\psi^{\LM}_\alpha H_a \\
\lambda^{\prime\beta} & \delta^\beta_\alpha(\tfrac 12-i D^\prime)-\sigma^{ab}{}_\alpha{}^\beta B_{ab} \\
\bar\lambda^{\prime\da} & 0\\
 D^\prime & (\DD_{\alpha\da}+\tfrac {3i}2H_{\alpha\da})\bar\lambda^{\prime\da}\\
 B_{ab}&-2i\DD_{[a}\bar\lambda^{\prime\da} \sigma_{b]\alpha\da}
+H_{[a}\sigma_{b]\alpha\da}\bar\lambda^{\prime\da}
-i\varepsilon_{abcd}H^c\sigma^d{}_{\alpha\da}\bar\lambda^{\prime\da} \\ H^a&\tfrac i2 \varepsilon^{abcd}\sigma_{b\alpha\da}\bar T_{cd}{}^\da\\
 \FF ab\dR&2\DD_{[a}\bar S^{\da} \sigma_{b]\alpha\da}
+iH_{[a}\sigma_{b]\alpha\da}\bar S^{\da}
+\varepsilon_{abcd}H^c\sigma^d{}_{\alpha\da}\bar S^{\da}\\
 \TT ab\beta&-i\delta_\alpha^\beta(\FF ab\dR+2\DD_{[a}H_{b]}) 
 -\tfrac 12\FF ab{cd}\sigma_{cd\alpha}{}^\beta
\\
 &
  -i\DD_{a}H^c\sigma_{bc\alpha}{}^\beta+i\DD_{b}H^c\sigma_{ac\alpha}{}^\beta
-3H^c H_{[a}\sigma_{bc]\alpha}{}^\beta\\
\bar T_{ab}{}^\da&0 \\
 \FF ab{cd}&-2\DD_{[a}\FF {b]}\alpha{cd}+2\TT \alpha{[a}\beta\FF {b]}\beta{cd}
+\bar T_{ab}{}^\da\FF \alpha\da{cd}\\
\multicolumn{2}{c}{}\\
\multicolumn{2}{c}{\mbox{Table 2}}
\ea
\]
$S^\alpha$ and $\bar S^\da$ are as in (\ref{S}) and are
the spin-1/2 parts of $\TT ab\alpha$ and $\bar T_{ab}{}^\da$:
\beq
S^\alpha=\TT ab\beta\sigma^{ab}{}_\beta{}^\alpha,\quad 
\bar S^\da=-\bar\sigma^{ab\da}{}_\dbe \bar T_{ab}{}^\dbe.
\label{S2}
\eeq
$\bar\DD_\da$ is obtained from $\DD_\alpha$ by complex conjugation, using
\[
\bar\DD_\da\bar{\cT}=(-)^{|\cT|}\,\overline{\DD_\alpha\cT}
\]
where $|\cT|$ denotes the Gra\ss mann parity of $\cT$.

The nonvanishing $\TT ABC$ and $\FF ABI$ in  (\ref{s5}) are $\TT ab\alpha$, $\TT ab\da=-\bar T_{ab}{}^\da$ and 
$\FF abI$ given in equations (\ref{Tab}) through (\ref{Fabi}), and the $\TT ABC$ and $\FF ABI$ given by
\bea
&&\TT \alpha\dbe c=\TT \dbe\alpha c = 2i\sigma^c_{\alpha\dbe}\label{TF1}\\
&&\TT \alpha b\gamma=-\TT b\alpha\gamma = -i\delta_\alpha^\gamma H_b+iH^c\sigma_{cb\alpha}{}^\gamma\label{TF2}\\
&&\TT \da b\dga=-\TT b\da\dga = i\delta_\da^\dga H_b+iH^c\bar\sigma_{cb}{}^\dga{}_\da\label{TF3}\\
&&\FF \alpha\dbe{cd}=\FF \dbe\alpha{cd}=2i \varepsilon^{abcd}\sigma_{a\alpha\dbe}H_b\label{TF4}\\
&&\FF \alpha b{cd}=-\FF b\alpha {cd}= -i\bar T^{cd\da}\sigma_{b\, \alpha\da}
       +2i\sigma^{[c}_{\alpha\da} \bar T^{d]}{}_b{}^\da\label{TF5}\\
&& \FF \da b{cd}=-\FF b\da{cd}=iT^{cd\alpha}\sigma_{b\, \alpha\da}
       -2i\sigma^{[c}_{\alpha\da} T^{d]}{}_b{}^\alpha\label{TF6}\\
&& \FF \alpha b\dR=-\FF b\alpha \dR=\sigma_{b\alpha\da}\bar S^\da\label{TF7}\\
&& \FF \da b\dR=-\FF b\da\dR=\sigma_{b\alpha\da} S^\alpha\label{TF8}\\
&& \FF \alpha bi=-\FF b\alpha i=-i\sigma_{b\alpha\da}\bar \lambda^{i\da}\label{TF9}\\
&& \FF \da bi=-\FF b\da i=i\sigma_{b\alpha\da} \lambda^{i\alpha}.\label{TF10}
\eea

The supercovariant derivative $\DD_a$ is defined on supercovariant tensors $\cT$ according to
\beq
\DD_a\cT=\Viel \mu a(\6_\mu-\tfrac 12\om_\mu{}^{ab}l_{ab}-A^\dR_\mu\delta_\dR-A_\mu^i\delta_i
-\psi_\mu^\alpha\DD_\alpha+\bar\psi_\mu^\da\bar\DD_\da)\cT.\label{Da}
\eeq
$\DD_a H^a$ and $\DD^a L^\LM_a$ fulfill the identities
\beq
\DD_a H^a=0,\quad \DD^a L^\LM_a=2H^a\DD_a\varphi^\LM-iS\psi^\LM+i\bar\psi^\LM\bar S.
\label{DHDL}
\eeq
$\DD_{[a}B_{cd]}$ and $H^a$ are related by
\beq
\DD_{[a}B_{bc]}=\tfrac 13\varepsilon_{abcd}H^d+i T_{[ab}\sigma_{c]}\bar\lambda^\prime
-i\lambda^\prime\sigma_{[a}\bar T_{bc]}.
\label{BH}
\eeq
For later purpose we remark that
$-iS$ plays the role of the gaugino of R-transformations, cf. equations (\ref{TF8}) and (\ref{TF10}), and that one has
\beq
\DD_\alpha S^\beta=\delta_\alpha^\beta (-\tfrac 14\FF ab{ba}+\tfrac 32 H^aH_a)
-i\FF ab\dR\sigma^{ab}{}_\alpha{}^\beta
\label{DS}
\eeq
which shows that $-\tfrac 14\FF ab{ba}+\tfrac 32 H^aH_a$ plays the role of the $D$-field of R-transformations.

The BRST transformation $s\cT$ of a supercovariant tensor $\cT$ is 
\beq
s\cT=(C^\mu\6_\mu+\xi^\alpha\DD_\alpha +\bar\xi^\da\bar\DD_\da+C^I\delta_I )\cT
=(\hat \xi^A\DD_A+\hat C^I\delta_I)\cT
\label{sten}
\eeq
wherein $C^i$ are ghosts of Yang-Mills gauge transformations
and $\hat\xi^A$ and $\hat C^I$ are ``covariant ghosts'' given by
\bea
&\hat\xi^a=C^\mu\viel a\mu ,\qd 
\hat\xi^\alpha=\xi^\alpha+C^\mu\psi_\mu^\alpha,\qd 
\hat\xi^\da=\bar\xi^\da-C^\mu\bar\psi_\mu^\da&\label{5xi}\\
&\hat C^{ab}=C^{ab}+C^\mu\om_\mu{}^{ab},\qd
\hat C^\dR=C^\dR+C^\mu A^\dR_\mu,\qd
\hat C^i=C^i+C^\mu A_\mu^i.&\label{5C}
\eea
The BRST transformations of $A_\mu^i$ and $A_{\mu\nu}^\LM$ are
\bea
sA_\mu^i&=&C^\nu\6_\nu A_\mu^i+(\6_\mu C^\nu)A_\nu^i
+\6_\mu C^i  +\f jki A_\mu^j C^k
-i\xi\sigma_\mu \bar\lambda^i+i\lambda^i\sigma_\mu \5\xi
\label{sAi}\\
sA_{\mu\nu}^\LM&=&C^\rho\6_\rho A_{\mu\nu}^\LM
+(\6_{\mu} C^\rho)A_{\rho\nu}^\LM+(\6_{\nu} C^\rho)A_{\mu\rho}^\LM
+\6_{\nu}Q_{\mu}^\LM-\6_{\mu}Q_{\nu}^\LM\nonumber\\
&&+2(\xi\sigma_{\mu\nu}\psi^\LM-\bar\psi^\LM\bar\sigma_{\mu\nu}\bar\xi)
+4i(\psi_{[\mu}\sigma_{\nu]}\bar\xi+\xi\sigma_{[\mu}\5\psi_{\nu]})\varphi^\LM
\label{sAM}
\eea
where $Q_\mu^\LM$ are real Gra\ss mann odd ghosts of reducible gauge transformations of $A_{\mu\nu}^\LM$. 
The BRST transformations of the ghosts $C^i$ and $Q_\mu^\LM$ are
\bea
sC^i&=&C^\mu\6_\mu C^i +\tfrac 12 \f jki C^k C^j
-2i \xi\sigma^\mu\5\xi\,A_\mu^i\label{sCi}\\
sQ_\mu^\LM&=&\6_\mu Q^\LM+(\6_\mu C^\nu)Q_\nu^\LM
-2i\xi\sigma^\nu\5\xi \, A_{\mu\nu}^\LM+2i\xi\sigma_\mu\5\xi\,\varphi^\LM\label{sQmuM}
\eea
where $Q^\LM$ are purely imaginary Gra\ss mann even ``ghosts for ghosts'' with ghost number 2 whose BRST transformations are
\beq
s Q^\LM=C^\mu\6_\mu Q^\LM-2i \xi\sigma^\mu\5\xi\,Q_\mu^\LM.\label{sQM}
\eeq
Covariant ghosts for ghosts $\hat Q^\LM$ are defined analogously to (\ref{5xi}) and (\ref{5C}) according to
\beq
\hat Q^\LM=Q^\LM+C^\mu Q_\mu^\LM +\tfrac 12 C^\mu C^\nu A_{\mu\nu}^\LM.
\label{5Q}
\eeq
The BRST transformations of the covariant ghosts and ghosts for ghosts are
\bea
s\hat\xi^A&=&-\tfrac 12(-)^{|B|} \hat\xi^B \hat\xi^C\TT CBA
+\hat C^I\Gg IBA \hat\xi^B                                                        \label{s5xi}\\
s \hat C^I  &=&
-\tfrac 12(-)^{|A|} \hat\xi^A \hat\xi^B\FF BAI
+\tfrac 12\f KJI \hat C^J \hat C^K                                                      \label{s5tC}\\
s\hat Q^\LM&=& \tfrac 16\hat\xi^a\hat\xi^b\hat\xi^c\varepsilon_{abcd}(L^{\LM d}-2H^d\varphi^\LM)
-2i\hat\xi^\alpha\hat\xi_{\alpha\da}\hat\xi^\da\,\varphi^\LM\nonumber\\
&&+\hat\xi^a\hat\xi^b(\hat\xi^\alpha\sigma_{ab\alpha}{}^\beta\psi_\beta^\LM
-\bar\psi_\da^\LM\bar\sigma_{ab}{}^\da{}_\dbe\hat\xi^\dbe)
\label{s5Q}
\eea
where $|a|=0$ and $|\alpha|=|\da|=1$.

The BRST transformations given above are strictly nilpotent off-shell. As is recapitulated in appendix \ref{app1},
the off-shell nilpotency of $s$ ($s^2=0$) on all fields (including the ghosts) 
except on $A_{\mu\nu}^\LM$, $Q_\mu^\LM$ and $Q^\LM$, and the construction of $\TT ab\alpha$, $\bar T_{ab}{}^\da$ and $\FF abI$ according to equations
(\ref{Tab}) through (\ref{Fabi}) can be deduced elegantly from the supercovariant
algebra (\ref{s5}) and the corresponding Bianchi identities. The nilpotency of $s$ on 
$A_{\mu\nu}^\LM$, $Q_\mu^\LM$ and $Q^\LM$ and the 
construction of $L_a^\LM$ according to equation (\ref{La}) can be checked separately.\footnote{In other words, one can check explicitly that $s$ squares to zero on $A_{\mu\nu}^\LM$, $Q_\mu^\LM$ and $Q^\LM$ and that 
the BRST transformation of $A_{\mu\nu}^\LM$ given in (\ref{sAM}) and the BRST transformations of $\varphi^\LM$ and $\psi^\LM$ arising from (\ref{sten}) imply that $L_a^\LM$ defined according to
equation (\ref{La}) transforms according to equation (\ref{sten}) with $\DD_\alpha L_a^\LM$ as in table 2. The BRST transformations of $A_{\mu\nu}^\LM$, $\varphi^\LM$ and $\psi^\LM$ and the definition of $L_a^\LM$ are compatible with equations (2.4) of \cite{Sohnius:1982fw}.} 
Furthermore the identities (\ref{DHDL}) can be checked explicitly.

\mysection{Invariant actions and elimination of auxiliary fields}\label{action}

The supercovariant algebra (\ref{s5}) and the way it is realized on the matter multiplets are exactly the same as in new minimal supergravity. In particular the additional fields 
$\lambda^\prime$ and $D^\prime$ do not occur in the supersymmetry transformations of
any component field of the matter multiplets, and the field $b_{\mu\nu}$ contributes to these supersymmetry transformations only via $H^\mu$ given in (\ref{Hb}), precisely as the 2-form gauge potential $t_{\mu\nu}$ in
new minimal supergravity. In addition the supercovariant algebra (\ref{s5}) is realized off-shell also on 
$\lambda^\prime$, $D^\prime$ and on the supercovariant tensor $B_{ab}$ given in (\ref{B}). For these reasons one can adopt methods and results derived in \cite{Sohnius:1982fw,Ferrara:1988qxa,Brandt:1993vd,Brandt:1996au} for new minimal supergravity to construct locally supersymmetric actions involving the matter multiplets and the fields $\lambda^\prime$ and $D^\prime$ in the present theory. 

In particular the results derived in \cite{Brandt:1993vd,Brandt:1996au} for invariant actions in new minimal supergravity
can be
extended straightforwardly to the presence of linear multiplets and of the additional supercovariant tensors 
$\lambda^\prime$, $D^\prime$ and $B_{ab}$. One obtains that Lagrangians which are invariant off-shell up to a total divergence, respectively, under the BRST transformations 
given in sections \ref{def} and \ref{trafos} are
\bea {\cal L}_{4}/e&=& (\bar\DD^2-4i\psi_\mu\sigma^\mu\bar\DD
+16\psi_\mu\sigma^{\mu\nu}\psi_\nu)
[\cA(\bar\phi,\bar\lambda,\bar\lambda^\prime,\bar S,\bar W)+\DD^2\cB({\cal T})]+c.c.\quad\label{L4}\\
{\cal L}_{5}/e&=&
\mu_{i_a}  (D^{i_a}
+\lambda^{i_a}\sigma^\mu\bar\psi_\mu+\psi_\mu\sigma^\mu\bar\lambda^{i_a}
+\varepsilon^{\mu\nu\rho\sigma}A^{i_a}_\mu \6_\nu b_{\rho\sigma})
\label{L5}\\
\cL_{6}/e&=&\kappa_{i_a\LM}[\tfrac 12\varepsilon^{\mu\nu\rho\sigma}A^{i_a}_\mu \6_\nu A_{\rho\sigma}^\LM
- \varphi^\LM D^{i_a}+(i\lambda^{i_a}\psi^\LM
-\varphi^\LM\lambda^{i_a}\sigma^\mu\bar\psi_\mu+c.c.)]\quad\label{L6}
\eea
where in (\ref{L5}) and (\ref{L6}) $\mu_{i_a}$ and $\kappa_{i_a\LM}$ denote real coupling constants and
the sum over $i_a$ runs over the abelian factors of $\cg$ including
the R-transformation with the identifications (cf. text around equation (\ref{DS}))
\beq
\lambda^\dR_\alpha\equiv -iS_\alpha,\quad
\5\lambda^\dR_\da\equiv i\bar S_\da,\quad
D^\dR\equiv -\tfrac 14\FF ab{ba}+\tfrac 32 H_aH^a.
\label{Rfields}
\eeq
In  (\ref{L4}) 
we used the notation
\beq
\bar\DD^2=\bar\DD_\da\bar\DD^\da,  \quad
\DD^2=\DD^\alpha\DD_\alpha,\quad
\bar W_{\da\dbe\dga}=-\bar T_{ab(\da}\bar\sigma^{ab}{}_{\dbe\dga)}
\eeq
and
$\cA(\bar\phi,\bar\lambda,\bar\lambda^\prime,\bar S,\bar W)$ denotes any function of the supercovariant tensors
$\bar\phi^m$, $\bar\lambda^i_\da$, $\bar\lambda^\prime_\da$, $\bar S_\da$ and $\bar W_{\da\dbe\dga}$ (but not of supercovariant derivatives thereof) which has
$R$-charge $-2$ and is invariant under all other generators $\delta_I$, and $\cB({\cal T})$ is any
funtion of supercovariant tensors which is invariant under all $\delta_I$,
\beq 
\delta_\dR\cA=-2i\cA,\quad
\forall I\neq\dR:\ \delta_I\cA=0,\quad \forall I:\ \delta_I\cB=0.
\label{act2}\eeq

$\cL_4$ is a generic Lagrangian which provides, amongst others, a standard locally supersymmetric Yang-Mills portion
arising from a contribution to $\cA$ proportional to $\bar \lambda^i\bar\lambda^jg_{ij}$ (with 
$\cg$-invariant metric $g_{ij}$), locally supersymmetric kinetic terms for the chiral multiplets arising from a contribution $b(\phi,\bar\phi)$ to $\cB$, 
superpotential terms for the chiral multiplets arising from a contribution $a(\bar\phi)$ to $\cA$ and locally supersymmetric kinetic terms for the linear multiplets arising from a contribution $c(\varphi)$ to $\cB$. In addition $\cL_4$ provides
Lagrangians with various higher derivative terms, such as four-derivative terms with the square of the Weyl tensor arising from a contribution
$\bar W\bar W$ to $\cA$ and/or with quartic terms in the Yang-Mills field strengths arising from a contribution to $\cB$ bilinear both in 
$\lambda$'s and $\bar\lambda$'s  (of course, $\cL_4$ provides further higher derivative terms, in particular terms with more than four derivatives). Furthermore,
contributions to $\cA$ given by $\tfrac 1{16}\bar\lambda^\prime\bar\lambda^\prime$ and 
$\tfrac i{8}\bar\lambda^\prime\bar\lambda^\prime$ reproduce $\cL^\prime_2$ and $\cL^\prime_3$ given in
(\ref{L2}) and (\ref{L3}), respectively.

$\cL_5$ and $\cL_6$ are ``exceptional Lagrangians'' that cannot be written in the form of $\cL_4$.
A contribution to $\cL_5$ with $i_a\neq \dR$ contains a Fayet-Iliopoulos term $e D^{i_a}$ and can thus
contribute to supersymmetry breaking \cite{Fayet:1974jb} and to the cosmological constant. 
The contribution to $\cL_5$ with $i_a= \dR$ reproduces for $\mu_\dR=-2M_{Pl}^2$ the 
$\cL_1$-portion of the Lagrangian 
(\ref{LSG}) as may be verified by explicitizing $\FF ab{ba}$, $S$ and $\bar S$. $\cL_6$ can contribute, amongst others, mass
terms for component fields of linear multiplets.

In appendix \ref{app2} solutions of the so-called descent equations are given 
which correspond to $\cL_4$, $\cL_5$ and $\cL_6$, respectively. 
It is easier to verify these solutions using reasoning given in \cite{Brandt:1993vd}  
than to check the invariance of $\cL_4$, $\cL_5$ and $\cL_6$ directly.

We now discuss the elimination of auxiliary fields for Lagrangians $\cL=\cL_4+\cL_5+\cL_6$ with $\cA$ and $\cB$ of the form
\beq
\cA=F(\bar\phi,\bar\lambda,\bar\lambda^\prime),\quad
\cB=G(\phi,\bar\phi,\varphi)=\overline{G(\phi,\bar\phi,\varphi)}.
\label{low}
\eeq
Such Lagrangians contain only terms with at most two derivatives and thus may be termed ``low energy Lagrangians''.
The fields $F^m$, $\bar F^{\bar m}$, $D^i$ and $D^\prime$ occur in $\cL$ undifferentiated and at most quadratically as can easily be checked. Hence, these fields can be eliminated algebraically using their equations of motion. However,
the direct algebraic elimination of the field $b_{\mu\nu}$ is hindered by terms in $\cL$ that are
quadratic in $H^\mu$. 
Such terms are present both in the contribution to $\cL_5$ with $i_a=\dR$ and (generically) in $\cL_4$ because $\cL_4$
contains, amongst others,
\bea
G_m\,\bar\DD^2\DD^2\phi^m=-2\,G_m\,\bar\DD^2F^m
=8i G_m\,(\tfrac 12\FF ab{ba}-3H^aH_a)\delta_\dR\phi^m+\ldots
\label{act3}
\eea
where 
\[
G_m=\frac {\6 G(\phi,\bar\phi,\varphi)}{\6\phi^m}\, .
\]
Now, one may remove the terms quadratic in 
$H^\mu$ by a suitable redefinition of the $R$-gauge field $A^\dR_\mu$.
To show this we collect all terms in $\cL$ containing $A^\dR A^\dR$, $A^\dR H$ or $HH$. 
A straightforward computation yields that these terms can be written as $\Delta\cL$ with\footnote{Here we used
$\delta_\dR G=G_m\delta_\dR\phi^m+G_{\bar m}\delta_\dR\bar\phi^{\bar m}=0$ where 
$G_{\bar m}=\6 G/\6 \bar\phi^{\bar m}$.\label{foot1}}
\beq
\Delta\cL/e=(\tfrac 32\mu_\dR-G_\dR)H_\mu H^\mu+G_{\dR\dR}A^\dR_\mu A^{\dR\mu}+2(\mu_\dR+G_\dR)A^\dR_\mu H^\mu
\label{Delta}
\eeq
where
\beq
G_\dR=48i G_m\, \delta_\dR\phi^m,\quad
G_{\dR\dR}=32\, G_{m\bar n}\, (\delta_\dR\phi^m)(\delta_\dR\bar\phi^{\bar n})
\label{Gs}
\eeq
with
\[
G_{m\bar n}=\frac {\6^2 G(\phi,\bar\phi,\varphi)}{\6\bar\phi^{\bar n}\6\phi^m}\, .
\]
We now make the following ansatz for a redefined R-gauge field:
\beq
A^{\prime\dR}_\mu=A^\dR_\mu+m H_\mu.
\label{Aprime}
\eeq
Using (\ref{Aprime}) in (\ref{Delta}) one obtains
\beq
\Delta\cL/e=u\, H_\mu H^\mu+G_{\dR\dR}A^{\prime\dR}_\mu A^{\prime\dR\mu}
+2(\mu_\dR+G_\dR-mG_{\dR\dR})A^{\prime\dR}_\mu H^\mu
\label{Delta2}
\eeq
with
\beq
u=m^2G_{\dR\dR}-2m(\mu_\dR+G_\dR)+\tfrac 32\mu_\dR-G_\dR.
\label{u}
\eeq
In order to remove all $HH$-terms from $\cL$ by the redefinition (\ref{Aprime}) $m$ has to be chosen such that $u$ vanishes. Obviously this can be achieved if 
$G_{\dR\dR}$ vanishes:
\beq
G_{\dR\dR}=0:\ u=0\ \Leftrightarrow\  m=\frac{3\mu_\dR-2G_\dR}{4(\mu_\dR+G_\dR)}\, .\label{case1}
\eeq
 However, $G_{\dR\dR}=0$ is a rather special case. For instance, $G_{\dR\dR}=0$ holds when all $\phi^m$ have vanishing $R$-charge which also implies $G_\dR=0$ and $m=\tfrac 34$ but forbids a superpotential. When $G_{\dR\dR}$ does not vanish, $u$ vanishes if
\beq
\left(m-\frac{\mu_\dR+G_\dR}{G_{\dR\dR}}\right)^2=
\frac{G_{\dR\dR}(G_\dR-\tfrac 32\mu_\dR)+(\mu_\dR+G_\dR)^2}{(G_{\dR\dR})^2}\, .\quad\label{case2}
\eeq
Now, $m$ must be real in order that $A^{\prime\dR}_\mu$ is real. In order to solve (\ref{case2}) with
real $m$ the numerator on the right hand side of equation (\ref{case2}) must not be negative.\footnote{$G_\dR$ and $G_{\dR\dR}$ are real. Indeed, $G_{\dR\dR}$  obviously is real because $G$ is real. Furthermore $\delta_\dR G=0$ implies (cf. footnote \ref{foot1})
$G_\dR=24i G_m\delta_\dR\phi^m
-24iG_{\bar m}\delta_\dR\bar\phi^{\bar m}
=24i G_m\delta_\dR\phi^m+c.c.$. For this reason $m$ in (\ref{case1}) is real too.} Whether or not this numerator is
non-negative depends on the $R$-charges of the $\phi^m$ and is not further discussed here. 

When $m$ is chosen such that 
$u$ vanishes and $A^{\prime\dR}_\mu$ instead of $A^\dR_\mu$ is used, the Lagrangian $\cL$ does not contain terms which are quadratic in derivatives of $b_{\mu\nu}$, and then $b_{\mu\nu}$ 
may be eliminated algebraically using the equations of motion if $F(\bar\phi,\bar\lambda,\bar\lambda^\prime)$ contains a term quadratic in $\bar\lambda^\prime$.

We also remark that $\tfrac 16(G_\dR-\tfrac 32\mu_\dR)$ is a generally field dependent prefactor of the Riemann curvature scalar in $\cL/e$,
cf. equation (\ref{act3}) (due to $\FF ab{ba}=R+\ldots$). This prefactor may be made field independent by a Weyl rescaling of the vierbein and corresponding redefinitions of other fields that convert the Lagrangian from Brans-Dicke form into conventional Einstein form, cf. \cite{Ferrara:1988qxa} for a detailed discussion of these field redefinitions in new minimal supergravity which analogously applies in our case. Furthermore we note that $G_{m\bar n}$ is the ``metric'' in the kinetic terms 
$eG_{m\bar n}\6_\mu\phi^m\6^\mu\bar\phi^{\bar n}$ of the $\phi^m$ and $\bar\phi^{\bar m}$ in $\cL$ (after integration by parts). If this metric is positive definite, $G_{\dR\dR}$ is non-negative.\footnote{However, according to \cite{Ferrara:1988qxa}, $G_{m\bar n}$ need not be positive definite in order that the kinetic terms for the $\phi^m$ and $\bar\phi^{\bar m}$ are positive after converting the Lagrangian into Einstein form.}

\mysection{Discussion}\label{con}

The formulation of D=4, N=1 supergravity studied in this paper is similiar to new minimal supergravity. 
This by itself is not surprising as this formulation was obtained as a consistent deformation of new minimal supergravity.
Nevertheless the deformation has some unusual and surprising features. 

One of these features is that in this formulation,
using the fields $b_{\mu\nu}$, $\lambda^\prime$ and $D^\prime$,
a Lagrangian, without or with matter fields included, in the simplest case differs from the corresponding Lagrangian of the
new minimal formulation of supergravity only by an added extra portion 
proportional to $\cL^\prime_2$ given in (\ref{L2}).\footnote{In
addition one may include $\cL^\prime_3$ given in (\ref{L3}) but this does not make much difference.} This extra portion
is separately invariant up to a total divergence under local supersymmetry transformations, and the remaining contributions to the Lagrangian are the same as in new minimal supergravity. 
The reason is that the fields $\lambda^\prime$ and $D^\prime$ which are not present in new minimal supergravity do not occur in the symmetry transformations of other fields except in the transformations of 
$\lambda^\prime$ and $D^\prime$ themselves and in the 
modified supersymmetry transformation of $b_{\mu\nu}$.
Furthermore, even though the symmetry transformations of $b_{\mu\nu}$ are modified as compared to new minimal supergravity, the symmetry transformations of $H^\mu$ defined in (\ref{Hb}) are not modified. Since the symmetry transformations of all other fields except $\lambda^\prime$ and $D^\prime$ depend on $b_{\mu\nu}$ 
at most via $H^\mu$ the modification of the symmetry transformations of $b_{\mu\nu}$ then has no impact on the  Lagrangian as compared to new minimal supergravity except for the added extra portion. 

In particular this implies that one may simply add to any Lagrangian of
new minimal supergravity, without or with matter fields included, the extra portion proportional to $\cL^\prime_2$. The resultant
theory again is an off-shell formulation of supergravity in which now $b_{\mu\nu}$ is a standard auxiliary field
without gauge degrees of freedom (which may be eliminated algebraically, at least for reasonable low energy Lagrangians, cf. section \ref{action}) and the R-gauge field is a physical field. Furthermore, 
if supersymmetry was unbroken before adding the extra portion, the inclusion of the extra portion 
inevitably introduces a cosmological constant, cf. equation (\ref{L2}), and breaks supersymmetry spontaneously 
(recall that $s\lambda^\prime=\tfrac 12\xi+\ldots$, cf. equation (\ref{sl}), i.e. $\lambda^\prime$ then is a goldstino that may be eaten by the gravitino). Thus, the addition of the extra portion particularly provides an alternative mechanism for spontaneously breaking local supersymmetry in new minimal supergravity, different from the familiar 
breaking mechanisms by Fayet-Iliopoulos terms or $F$-terms.

On the other hand, in addition to or in place of $\cL^\prime_2$ given in (\ref{L2}) and optionally $\cL^\prime_3$ given in (\ref{L3}), one may include other terms in the Lagrangian depending on 
$b_{\mu\nu}$, $\lambda^\prime$ and $D^\prime$, such as terms arising from contributions to $\cA$ in (\ref{low}) which depend on 
both $\bar\lambda^\prime$ and some $\bar\phi^{\bar m}$. This may have a more subtle effect on the theory as compared to new minimal supergravity and may be worth a further study.

\appendix

\mysection{BRST approach to off-shell supergravity}\label{app1}

In this appendix we recapitulate briefly a BRST approach to off-shell supergravity theories which was used also in \cite{Brandt:1993vd,Brandt:1996au} and applies a general framework \cite{Brandt:1996mh} of treating theories with
local and/or global symmetries. The approach is based on a supercovariant
algebra (\ref{s5}) and corresponding Bianchi identities
\bea & &\csum {ABC}{36}(-)^{|A|\, |C|}
     (\DD_A\TT BCD+\TT ABE\TT ECD+\FF ABI\Gg ICD)=0
                                                          \label{A1}\\
 & &    \csum {ABC}{36}(-)^{|A|\, |C|}
     (\DD_A\FF BCI+\TT ABD\FF DCI)=0
                                                          \label{A2}\eea
where ${\displaystyle \csum {}{25}}X_{ABC}=X_{ABC}+X_{BCA}+X_{CAB}$ denotes the cyclic sum.

The supercovariant
algebra (\ref{s5}) and the Bianchi identities (\ref{A1}), (\ref{A2}) are supposed to be realized off-shell on supercovariant tensor fields. The approach uses a differential which is the sum of the BRST differential $s$ and the exterior
derivative $d=dx^\mu\6_\mu$ and is denoted by $\4 s$,
\beq
\4 s = s+d.
\label{A3}
\eeq
This differential acts on total forms and has total degree 1. A total form generally is a sum of local $p$-forms with
various form degrees $p$. The total degree is the sum of the ghost number and the form degree. Hence, a total form
$\4\omega^g$ with definite total degree $g$ (``total $g$-form'') generally is a sum 
$\4\omega^g=\sum_p \omega^{p,g-p}$ of
local $p$-forms $\omega^{p,g-p}$ where $\omega^{p,g-p}$ has ghost number $g-p$.

The generators $\DD_A$ and $\delta_I$ of the
algebra (\ref{s5}) are related to total 1-forms constructed of the ghosts and corresponding 1-forms according to
\bea
\4\xi^a&=&\hat\xi^a+dx^\mu\viel a\mu=(C^\mu+dx^\mu)\viel a\mu\label{A4}\\
\4\xi^\alpha&=&\hat\xi^\alpha+dx^\mu\RS \alpha\mu=\xi^\alpha+(C^\mu+dx^\mu)\RS \alpha\mu\label{A5}\\
\4\xi^\da&=&\hat\xi^\da-dx^\mu\bRS \da\mu=\5\xi^\da-(C^\mu+dx^\mu)\bRS \da\mu\label{A6}\\
\4C^I&=&\hat C^I+dx^\mu A^I_\mu=C^I+(C^\mu+dx^\mu)A^I_\mu\label{A7}
\eea
where in the present case $\{A^I_\mu\}=\{\omega_\mu{}^{ab},A^\dR_\mu,A^i_\mu\}$.
$\4 s$ acts on supercovariant tensors $\cT$ and on the total 1-forms $\4\xi^A$ and $\4 C^I$ according to
\bea
\4 s\,\cT&=&(\4\xi^A\DD_A+\4 C^I\delta_I)\,\cT\label{A8}\\
\4s\, \4\xi^A  &=&-\tfrac 12(-)^{|B|} \4\xi^B \4\xi^C\TT CBA
+\4C^I\Gg IBA \4\xi^B                                                       \label{A9} \\
\4s\, \4C^I  &=&
-\tfrac 12(-)^{|A|} \4\xi^A \4\xi^B\FF BAI
+\tfrac 12\f KJI \4C^J \4C^K.                                                    \label{A10}
\eea
Using the algebra (\ref{s5}) and the Bianchi identities (\ref{A1}), (\ref{A2}) one easily
checks that $\4 s$ squares to zero on supercovariant tensors $\cT$ and on the total 1-forms $\4\xi^A$ and $\4 C^I$.

The ghost number 0 part of (\ref{A8}) provides the supercovariant derivatives $\DD_a$ of supercovariant tensors as it
yields 
\[ \6_\mu\cT=(\viel a\mu\DD_a+\RS \alpha\mu\DD_\alpha-\bRS \da\mu\bar\DD_\da+A_\mu^I\delta_I)\cT\]
which can be solved for $\DD_a\cT$ and then gives equation (\ref{Da}) in the present case. 
The ghost number 1 part of
(\ref{A8}) provides the BRST transformations (\ref{sten}) of supercovariant tensors.

The ghost number 0 parts of (\ref{A9}) for $A=\alpha$ and  $A=\da$ and of (\ref{A10}) provide the 
field strengths (or curvatures)
$\TT ab\alpha$, $\TT ab\da$ and $\FF abI$, respectively. For instance, the ghost number 0 part of
(\ref{A9}) yields for $A=\alpha$
\[
\6_\mu\RS \alpha\nu-\6_\nu\RS \alpha\mu=\viel a\mu\viel b\nu\TT ab\alpha+\ldots
\]
which can be solved for $\TT ab\alpha$ and then gives (\ref{Tab}) in the present case. Analogously
one obtains equations (\ref{barTab}) through (\ref{Fabi}) from (\ref{A9}) and (\ref{A10}). The ghost number 0 part of
 (\ref{A9}) for $A=a$ provides analogously either $\TT bca$, or it determines $\omega_\mu{}^{ab}$ when the constraint $\TT bca=0$ is imposed. In the present case this gives $\omega_\mu{}^{ab}$ as in
equations (\ref{spin1}) and  (\ref{spin2}). 

The ghost number 1 parts of (\ref{A9}) and of (\ref{A10}) provide the
BRST transformations of $\viel a\mu$, $\RS \alpha\mu$ and $A^I_\mu$. In the present case one obtains in particular
equations (\ref{sviel}), (\ref{spsi}), (\ref{sa}) and (\ref{sAi}). The ghost number 2 parts of (\ref{A9}) and of (\ref{A10})
then provide the BRST transformations of the ghosts $C^\mu$, $\xi^\alpha$ and $C^I$. In the present case this
gives equations (\ref{s5xi}) and (\ref{s5tC}) and then, using in addition the BRST transformations of $\viel a\mu$, $\RS \alpha\mu$ and $A^I_\mu$, equations (\ref{sCdiff}) through (\ref{sc}) and equation (\ref{sCi}).

As we remarked at the end of section \ref{trafos}, the BRST transformations of
$A_{\mu\nu}^\LM$, $Q_\mu^\LM$ and $Q^\LM$ given in equations (\ref{sAM}), (\ref{sQmuM}) and (\ref{sQM})
and the field strength $L^\LM_a$ given in equation (\ref{La}) cannot be deduced from the algebra
(\ref{s5}) in the same manner. Nevertheless these equations can also be written in a compact form which is analogous to
(\ref{A9}) and of (\ref{A10}) and reads
\bea
\4 s \,\4 Q^\LM&=&\tfrac 16\4\xi^a\4\xi^b\4\xi^c\varepsilon_{abcd}(L^{\LM d}-2H^d\varphi^\LM)
-2i\4\xi^a\4\xi^\alpha\sigma_{a\alpha\da}\4\xi^\da\,\varphi^\LM\nonumber\\
&&+\4\xi^a\4\xi^b(\4\xi^\alpha\sigma_{ab\alpha}{}^\beta\psi_\beta^\LM
-\bar\psi_\da^\LM\bar\sigma_{ab}{}^\da{}_\dbe\4\xi^\dbe)
\label{A11}
\eea
where
\bea
\4 Q^\LM&=&\hat Q^\LM+dx^\mu Q_\mu^\LM +\tfrac 12 dx^\mu dx^\nu A_{\mu\nu}^\LM\nonumber\\
&=&Q^\LM+(C^\mu+dx^\mu) Q_\mu^\LM +\tfrac 12 (C^\mu+dx^\mu) (C^\nu+dx^\nu) A_{\mu\nu}^\LM.
\label{A12}
\eea

\mysection{Solutions of the descent equations}\label{app2}

In this appendix we provide solutions of the so-called descent equations that contain the Lagrangians
$\cL_4$, $\cL_5$ and $\cL_6$ given in equations
(\ref{L4}), (\ref{L5}) and (\ref{L6}), respectively. These descent equations read
\beq 0<p\leq 4:\qd s\om_i^{p,4-p}+d\om_i^{p-1,5-p}=0,\qd
                   s\om_i^{0,4}=0                                                         \label{A13}\eeq
where $\om_i^{p,4-p}$ is a local $p$-form with ghost number $4-p$. The 4-form 
$\om_i^{4,0}=\cL_i d^4x$ contains the respective Lagrangian $\cL_i$ for $i=4,5,6$. The descent equations
can be written compactly by means of total 4-forms $\4\omega_i$ as
\beq
\4 s\,\4 \omega_i=0,\quad\4\omega_i=\sum_{p=0}^4\om_i^{p,4-p}.
\label{A14}
\eeq

Total 4-forms $\4\omega_4$ and $\4\omega_5$ containing the Lagrangians $\cL_4$ and $\cL_5$, respectively,
can be obtained from \cite{Brandt:1993vd}. They can be written as
\bea
\4\omega_4&=&(-4i\bar\vartheta\bar\vartheta-4i\bar\eta\,\bar\DD+\Xi\bar\,\bar\DD^2)
[\cA(\bar\phi,\bar\lambda,\bar\lambda^\prime,\bar S,\bar W)+\DD^2\cB({\cal T})]+c.c.
\label{A15}\\
\4\omega_5&=&\mu_{i_a}(2\4C^{i_a}\4 H
+\bar\lambda^{i_a}\bar\eta+\eta\lambda^{i_a}+\Xi \, D^{i_a})
\label{A16}
\eea
where
\bea
&\bar\vartheta^\da=\4\xi^{\da\alpha}\4\xi_\alpha,\
\eta^\alpha=-\tfrac i{6}\,\vartheta^\beta\4\xi_{\beta\da}\4\xi^{\da\alpha},\
\bar\eta^\da=\tfrac i{6}\,\4\xi^{\da\alpha}\4\xi_{\alpha\dbe}\bar\vartheta^\dbe,\
\Xi=-\tfrac 1{24}\, \varepsilon_{abcd}\, \4\xi^a\4\xi^b\4\xi^c\4\xi^d&\quad\label{A17}\\
&\4 H=\tfrac 16\,\4\xi^a \4\xi^b \4\xi^c\varepsilon_{abcd}H^d+
      i\4\xi^\alpha\4\xi_{\alpha\da}\4\xi^\da.&
                                                              \label{A18}
\eea
Using reasoning given in \cite{Brandt:1993vd} it is
actually not difficult to verify that $\4\omega_4$ and $\4\omega_5$ fulfill (\ref{A14}). 
A total 4-form $\4\omega_6$ which contains the Lagrangian $\cL_6$
is
\beq
\4\omega_6= \kappa_{i_a\LM}[\4C^{i_a}\4s\4Q^\LM
-(\eta\lambda^{i_a}+\bar\lambda^{i_a}\bar\eta)\varphi^\LM
+\Xi(-D^{i_a}\varphi^\LM+i\lambda^{i_a}\psi^\LM-i\bar\psi^\LM\bar\lambda^{i_a})] \label{A19}
\eeq
with $\4s\4Q^\LM$ as in (\ref{A11}). $\4s\4\omega_6=0$ can be shown similarly to $\4s\4\omega_5=0$
in \cite{Brandt:1993vd}.

\end{document}